\let\oldmarginpar\marginpar
\renewcommand\marginpar[1]{\-\oldmarginpar[\raggedleft\footnotesize #1]%
{\raggedright\footnotesize #1}}

\documentclass[english,11pt]{article}
\usepackage[left=2cm,top=1cm,right=6cm,nohead,nofoot]{geometry}
\newcommand{\ess}{\lambda}

\usepackage{graphicx}
\usepackage{listings}
\usepackage{babel}
\usepackage{ucs}
\usepackage[utf8x]{inputenc}

\newtheorem{expl}{Description}[section]

\newcommand{\set}[1]{\{#1\}}

\newcommand{\Word}{\mathrm{Word}}

\newcommand{\Addr}{\mathrm{Addr}}
\newcommand{\Reg}{\mathrm{Reg}}
\newcommand{\IB}{\mathrm{InstructionBoundary}}
\newcommand{\Active}{\mathrm{Active}}

\newcommand{\Bfunc}{\left\{\begin{array}{ll}}
\newcommand{\Efunc}{\end{array}\right.}

\newcommand{\Beq}{\begin{eqnarray}}
\newcommand{\Eeq}{\end{eqnarray}}

\usepackage[T1]{fontenc}
\usepackage{cmbright}
\usepackage[bookmarks=false,colorlinks=true,citecolor=black]{hyperref}
\title{ The meaning of concurrent programs (DRAFT)} 
\author{Victor Yodaiken\\
{\small Copyright 2008.\footnote{Permission granted to make and distribute
complete copies for non-commercial use but not for use in a publication. All other
rights reserved but fair use encouraged as long as properly cited.}}\\
\url{yodaiken@finitestateresearch.com}}
\begin{document}
\maketitle

\section{Basics}

Consider a combined software and hardware "system" consisting of a set of threads $T$, a memory, a set
of devices $D$, some 
number of processor cores, and i/o and other components that we don't need to specify yet. 
Questions like "can there be a state where multiple threads are executing inside a critical region"
can be answered only
by understanding how state variables change as the system changes. Since the hardware and software are
designed to be both discrete state and deterministic
we can
consider system variables to be \emph{functions of the sequence of events that have driven the system to
its current state.} For example, \emph{the contents} stored in a memory location
vary with the event sequence. Or consider the following: 
\marginpar{
"Formal methods" researchers argue concurrency is non-deterministic because
they have confused "unspecified" with "non-deterministic".}
 \begin{quote}{thread $t$ starts a test-and-set operation
with arguments "ptr" $=\alpha$ and "index" $=j$ in the state determined by event sequence $w$}\end{quote}
The event sequence is generally immense and the events are complicated. A single event may
correspond to the signal changes on the input pins of every circuit on the system during a single processor
cycle. But we can abstract out properties of the sequences and focus on the
properties of interest.

The "specifications" given here are in ordinary working mathematical notation plus some relatively informal
language. I have made a deliberate effort to try to avoid unnecessary formalization. A statement
of the form "at this state "x" holds contents $=j$" is clear enough --- and can be translated into more formal
mathematical notation whenever needed. See section \ref{sec:details} for details.
On the other hand, I have made efforts to avoid oversimplifying the 
semantics of actual computations. For example, the execution of an atomic "compare and swap" operation is
both impressively complex and precise. When a thread reaches the start of this operation, there may be interrupts
which cause unspecified delays as the operating system switches out the thread and any number of other tasks
may start the same operation "at the same time", but the hardware assures that only one thread will complete
the operation and get a success result. 

\subsection{Sequences and state}
Given a sequence $w$ and a variable depending on $w$, $x$ the meaning of 
\emph{$x$ at $w$} is simply the value of $x$ in the state reached by following
$w$ from the initial state. For example consider:
$$\mbox{The contents of  memory location}\alpha \mbox{ is }n\mbox{ at } w  .$$

Many properties are described in terms of what
can happen between two states. Write $wz$ for the sequence obtained by appending
sequence $z$ to sequence $w$. 
If some memory location $\alpha$ remains
unchanged from the $w$ state to the $wz$ state we could say:
\begin{quote}{\em
The contents of location $\alpha$ does not change between $w$ and $wz$ (inclusive of the end points).
}\end{quote}
I'll label memory locations with either addresses or symbolic names that
may depend on the thread. The address corresponding to variable {\tt x} for
task $t$ at $w$ may not be the same as the address of {\tt x} for thread $t'$
or even for thread $t$ in another state. Write $\Addr(w,t,\mathtt{x})$ for the
address of {\tt x} in the state determined by $w$ in the context of $t$.
Let $\Word(w,\alpha)$ be the contents of memory at address $\alpha$ in the $w$ state. 

Let's insist that memory contents only change if a device or thread "writes" to
that location:
\begin{expl}
If $\Word(w,\alpha)\neq \Word(wz,\alpha)$ there there must be some 
be some thread $t$ so that $t$ writes $\Word(wz,\alpha)$ to $\alpha$ between $w$ and $wz$
or some device $d$ so that $d$ writes $\Word(wz,\alpha)$ to $\alpha$ between $w$ and $wz$
\end{expl}

When we discuss the way a state variable changes as the system changes state,
we will often need to be able to identify the states that are visited 
in between the two terminal states. For any $u$, the
sequence $w$ is a {\em prefix} of the sequence $wu$. If $u$ is not the empty
sequence, then $w$ is a {\em proper} prefix of $wu$. Write $w\leq wz$ to 
indicate $w$ is a prefix, and $w< wu$ to indicate that $w$ must be a proper
prefix. Then $w < q < wz$ constrains $q$ to be between $w$ and $wz$ and but
not equal to either, while $w \leq q \leq wz$ allows $q$ to reach the end points. So our constraint above could be rewritten more precisely as

\begin{expl}
If $\Word(w,\alpha)\neq \Word(wz,\alpha)=k$ there there must be some 
$w< q\leq wz$ so that 
for some thread $t$, $t$ writes $k$ to $\alpha$ at $q$ or 
or for some device $d$, $d$ writes $k$ to $\alpha$ at $q$.
\end{expl}

The problem with shared memory systems is that the operation of reading,
modifying and writing a new value back is generally not atomic. This 
can be shown as follows for a simple increment {\tt x = x+1} the
semantics of which is as follows.

\begin{expl}
If task $t$ starts {\tt x = x + 1} at $w$ and completes it at $wz$ and
$\Addr(w,t,x)=\alpha$ then there is some $w\leq q \leq wz$ so 
that $\Word(wz,\alpha)=\Word(q,\alpha)+1$
(where "+1" depends  on the type of the variable
{\tt x} in the context of $t$).
\end{expl}

Suppose $t$ executes {\tt x = x+1} in the interval $w$ to $wz$ and
$t'$ executes the line of code in the interval $w'$ to $wz'$ where 
$w \leq w' < wz \leq wz'$ and for $t$ we have
$\Word(wz,\alpha)=\Word(q,\alpha)+1$  and for $t'$ we have
$\Word(wz',\alpha)=\Word(q',\alpha)+1$. There is no assurance that
$\Word(q',\alpha) = \Word(wz,\alpha)$ --- and that's the heart of the
synchronization problem for data in shared memory architectures.

An atomic
compare and swap (ACS) operation changes swaps the contents of memory for a "new" value if it finds the contents to be identical to a "test" value --- {\em and } if 
there is no competing write that beats us to the punch. If the
result is $1$ then (1) no other task can get a result of $1$ for that
address during the interval, and (2) at the start, $\alpha$ contains the
old value and it only changes in some in-between state to contain the new
value. If the result is $0$ then (1) the thread does not complete any
write operation during the interval, and (2) there is, by way of
explanation, some in-between state where $\alpha$ does not 
contain the expected old value or some intervening "write" operation.
\begin{expl}
If thread $t$ starts an atomic compare-and-swap (ACS) operation at $w$  with target$=\alpha$, new$=n$ and old$=k$ and completes it at $wz$, then
let $R_{t, wz,\alpha}=$ the result at $wz$ for $t$
\Beq
R_{t,wz,\alpha}\in\set{0,1}\\
R_{t,wz,\alpha}=1\rightarrow \mbox{ for all }w\leq q\leq wz\mbox{ there is no }t'\neq t, R_{t',q,\alpha}=1\nonumber\\
\mbox{ and there is some }w< q\leq wz\mbox{ so that } \nonumber\\
\mbox{for all }w\leq q'< q\mbox{ the contents of }\alpha\mbox{ at }q'\mbox{ is }k\nonumber\\
\mbox{and for all }q\leq q'\leq wz\mbox{ the contents of }\alpha\mbox{ at }q'\mbox{ is }n\\
R_{t,wz,\alpha}=0\rightarrow \mbox{ for all  }w\leq q\leq wz,t\mbox{ does not write to memory at }q\nonumber\\
\mbox{ and there is some }w\leq q\leq wz\mbox{ so that either} \\
\mbox{ the contents of }\alpha\mbox{ at }q\mbox{ is not }k\nonumber\\
\mbox{ or there is some write at }q \mbox{ by any device or a thread }t'\neq t
\Eeq
\end{expl}

Note that we do not require the hardware is smart enough to be sure 
that we succeed if some in-between write writes the old value $k$. This
allows for implementation by hardware that does a "clear written bit on
this address", then a "load contents", then a "write if written bit is still
zero". And there is no requirement that the ACS complete in any fixed time - that's something we'd need in a more detailed treatment.

\subsection{Pointers, functions, and longer chunks of code}
$\Word(w,\Word(w,\alpha))$ is the contents of the memory at the address that
is the contents of the memory at $\alpha$ in the $w$ state. Consider this simple function.

\lstset{language=c}
\begin{lstlisting}
void calculate(int m, int *ptr){
	int old = *ptr;
	*ptr = m*m + *ptr;	
	return old;
}
\end{lstlisting}

The intended behavior can be defined as follows:

\begin{expl}
If $t$ starts to call $\mathtt{calculate}$ with "m" $=j$ and "ptr" $=\alpha$  at $w$\\
and $t$ returns from the call started at $w$ in $wz$.\\
Then $\Word(wz,\alpha)=\Word(w,\alpha)+j*j$ and at $wz$ the return value of $t$ $=\Word(w,\alpha)$. (Assuming non-interference).
\end{expl}

What's non-interference? In this case it is just that:
$$\mbox{There is no }w\leq q\leq wz,h\in T\setminus\set{t}\cup D,\mbox{ so that }h\mbox{ writes to any of the local variables of }t\mbox{ at }q$$

\subsubsection{Note on machine model}
The model used here assumes that "writes" commit at the last event --- so that
a store to memory location $\alpha$ may takes multiple events, but $\Word(w,\alpha)$ only changes as the write completes. I can't see how this assumption conflicts with computer architecture practice in any way that would lead us astray, but
the assumption is not at all necessary for using the methods described here.

More seriously, I'm glossing over non-coherent memory here just to simplify exposition. In fact, $\Word(w,t,\alpha)$ may not equal $\Word(w,\alpha)$ if some $t'$
has written to $\alpha$ but the new value is in a write buffer or even if the
write has been executed out of order. I'll return to this below to show how to
make the model more realistic, but assuming
that memory is coherent is reasonable in many situations and leaves us with 
a useful model.

I'm treating memory contents as "numbers" --- assuming that expressions like
$\Word(w,\alpha)+1$ are known to be shorthand for e.q.
$\Word(w,\alpha)+1)\bmod 2^{32}$ or whatever the programming language type
restrictions call for.
Finally, I'm only working with whole words of memory value here and am not
worrying about bytes --- see section \ref{sec:details} for some discussion.

\section{Critical regions}

One protocol for synchronization is to use a memory location as a "gateway" set to contain $0$ when open and some non-zero value, say $1$, when closed. 
Once the gateway is initialized, we can require that threads succeed in an
atomic compare and swap with the gateway address as target, $0$ as the old
value, and $1$ as the new value to become "owner" and that the gateway is
released by setting it to zero. It's not necessary to have the owner always
be the releaser - but the releaser needs to be sure not to release an already
released gateway.
To understand this problem, suppose $t_1$ is trying to enter
the gateway and $t_2$ is trying to release it --- but it is already released.
Then $t_1$ may fail on the ACS operation because a write happens during
the ACS operation --- even though the write does not change the contents.

I'm going to define $G(w,\alpha)\in\set{0,1}$ to tell us if the 
gateway has been initialized and used properly and then $Owns(w,\alpha,t)\set{0,1}$ to tell us if thread $t$ owns the closed gateway.
Let's leave "activated" and "deactivated" undefined for
now and just track status. The empty sequence of events "$\ess$" is the 
sequence that leads to the initial state. So if we define a function at $\ess$
and at $wa$ in terms of its value at $w$, we have defined it for every
state.
\Beq
G(\ess,\alpha)=0\\
G(wa,\alpha) = \Bfunc
1&\mbox{if the gateway is set to }0\mbox{ and was activated}\\
&\mbox{ and no thread is executing an ACS operations with target=}\alpha\\
0&\mbox{if the gateway was deactivated}\\
&\mbox{or if some device }d\mbox{ writes to}\alpha\mbox{ at }wa\\
&\mbox{or if some thread }t\mbox{ writes a nonzero value to }\alpha\mbox{ at }wa\\
&{    unless }t\mbox{ is executing an ACS operation }\\
&\mbox{or if some thread }t\mbox{ writes a zero value to }\alpha\mbox{ at }wa\\
&\mbox{    unless }\alpha\mbox{ contains }1\mbox{ at }w\\
G(w)&mbox{otherwise}
\Efunc\\
Owns(\ess,\alpha,t)=0\\
Owns(wa,\alpha,t) = \Bfunc
	0&\mbox{if }G(wa,\alpha)=0\\
	&\mbox{or if }\alpha\mbox{ contains }0\mbox{ at }wa\\
	1&\mbox{if }G(wa,\alpha)=1\\
	&\mbox{and }t\mbox{ completes an ACS operation}\\
	&\mbox{with target=}\alpha,\mbox{old=}0\mbox{ new=}1\mbox{ and result=}1\mbox{ at }wa
\Efunc
\Eeq

We can now show that:
\Beq \Sigma_t Owns(w,\alpha,t) \leq 1\Eeq
This is obviously correct if $G(w,\alpha)=0$, so in what follows assume
$G(w,\alpha)=1$. 
\Beq \Sigma_tOwns(w,t,\alpha)\leq 1\mbox{ and }\Sigma_t Owner(w,t,\alpha)>0 \leftrightarrow \Word(w,\alpha)=1\Eeq

[Proof is done, but ugly. Basic idea is induction on string length. TBFixed].


Let $C$ be a set of line numbers within a "critical region". We may want to use ACS operations to 
guard a critical operation.
So we may want to show that for some $\alpha$
\Beq \mbox{if }t \mbox{ is executing a line }n\in C\mbox{ in the } w\mbox{ state  then }Owner(w,t,\alpha).\Eeq

\section{Details\label{sec:details}}
Assume we have $\Word$ and also $\Reg$ so that $\Reg(w,t,r)$ is the contents
of either the physical register $r$ in the $w$ state if $t$ is executing on
some core in that state, or the stored register saved by the OS if $t$ is blocked
in that state. We also need $\IB(w,c)\in\set{0,1}$ to be true (1) if and only if core $c$
completes execution of its current instruction in the $w$ state.  Finally, 
we need some understanding of how the OS tracks threads - let
$\Active(w,c,t)\in\set{0,1}$ be true (1) if and only if thread $t$ is executing
on core $c$ in the $w$ state. In most operating systems, there will be a 
data structure indexed by core processor identifier so that we will
have something like
\Beq
\Active(w,c,t)= \Bfunc
	1&\mbox{if } \Word(w,\beta) = t, \mbox{ where }\beta = \Word(w,\Word(w,"current")+c)\\
	0&\mbox{otherwise}
\Active(w,t) = \Sigma_c \Active(w,c,t)
\Efunc \Eeq

\begin{description}
\item[] Assume that $Active(w,t) \leq 1$.
\item[] For each thread $t$, $t$ is executing  at $w$ if and only if $\Active(w,t)$.
\item[] Thread $t$ writes value $j$ to memory location $\alpha$ at $w$ depends
on $\Reg(w,t,programcounter)$  and $\IB(w,c)$ where $c$ is the core
identifier so that $\Active(w,c,t)$. If $\Active(w,t)=0$ then the thread cannot
be completing a write at $w$.
\item[] Thread $t$ starts execution of line of code \tt{y = x+1} at $w$ and 
completes execution of the line of code at $wz$ also depends on $\Reg$ and
$\IB$. 
\item[] Thread $t$ calls function $\mathtt{f(int\ x, float \ y)}$ with arguments "x" $=i$, "y" $=j$ at $w$ and completes the call with return value $=k$ at
$wz$ requires some depth tracking if we permit recursive functions --- which we should. Let $Fdepth(\ess,t,f)=0$ and 
$$Fdepth(wa,t,f)= \Bfunc
	1+Fdepth(w,t,f)&\mbox{if }t\mbox{ calls }f\mbox{ at }wa\\
	Fdepth(w,t,f)-1&\mbox{if }t\mbox{ ends a call to }f\mbox{ at }wa\\
	Fdepth(w,t,f)&\mbox{otherwise}\Efunc
$$
Then $t$ calls {\tt f} at $w$ and returns from that call at $wz$ requires that
$Fdepth(w,t,f) = Fdepth(wz,t,f)+1$ and there is no $w < q < wz$ so that
$Fdepth(w,t,f) = Fdepth(q,t,f)+1$.
\end{description}
\section{Related Work and Empiricism versus Axiomatics}

This is a less formal and less OS-centric companion to \cite{h2} which is a successor to a long series of papers attempting to
make this line of research into something practical. 

This work is in some ways a reaction against the entire field of "formal methods
" which starts with the idea that a program is a mathematical object that
can and should be "formalized". I'm more comfortable with considering a 
program to be a manufactured object with some properties we may find useful to define mathematically but with a nature that is empirical. So my goal is to 
provide methods that can be used in conjunction with  informal rules, and experimentation, and testing, much engineers approach other manufactured goods such as locomotives and rubber ducks.

The empirical bias lead me to discard the emphasis on non-determinism in 
the formal methods literature. In software and hardware design, non-determinism
is an error condition or is a result of interaction with some partially
specified device or software component. At the most basic, if we see
systems as non-deterministic, they must be modelled as relations: a sequence
of events $w$ maps to a set of possible terminal states. But relations are
really awkward objects and it is conceptually at least as reasonable to 
consider each sequence to determine a single terminal state --- but one which 
we may not be able to fully specify. Even the most non-deterministic of 
phenomena, such as a gate that can go into meta-stable state can be considered
a deterministic device. Is the state machine that models the gate 
non-deterministically choosing an output or reading from a very large or even infinite table of random digits? I can't see why we would ever care at the 
system level.

The techniques of formal logic/meta-mathematics and the viewpoint rooted in
the semantics of programming languages have drawbacks for a more empirical
approach to semantics.  Applied mathematicians do not use formal logic
- formal logic is a tool for reasoning about mathematics while I'm more
interested
in reasoning about test-and-set bit instructions.  And programming
languages, especially those which have built-in "concurrency" have weak
semantics that requires building up of complex rule sets.
For example, the treatment of concurrent threads here is far
simpler than that of Milner\cite{Milner} and Hoare \cite{Hoare} where a thread
has to be treated as a fundamental object that is inherently "non-deterministic"
instead of as product of an underlying deterministic scheduling system.

It may be obvious, however, that the ideas of reasoning about intervals
were influenced and derive a great deal from works on temporal
logic\cite{Manna,Moszkowski} and more generally modal logics\cite{Kripke}. The idea of dealing with sequences of events
instead of states comes from frustrating attempts to describe specific 
paths using the state quantifiers in temporal logic. Temporal logic
allows the user to say "P is true in the all possible next states" or 
"P is true in some possible next states", but to say "if X happens and drives
us to the next state, then P" requires additional data structures and after
some one one begins to doubt the utility of the formal logic framework.

\bibliography{all}
\bibliographystyle{alpha}
\end{document}